# An Iterative Method for Calculating the Polar Component of the Molecular Solvation Gibbs Energy Under a Smooth Change in the Dielectric Permittivity of a Solution

Grigoriev* F.V., Kupervasser ** O.Yu., Kikot' *** I.P.

*E-mail:* fedor.grigoriev@gmail.com,
**E-mail:* olegkup@yahoo.com,
***E-mail:* irakikotx@gmail.com

An iterative method for calculating the polar component of the solvation Gibbs energy under a smooth change in dielectric permittivity, both between a substrate and a solvent and in a solvent is formulated on the basis of a previously developed model. The method is developed in the approximation of the local relationship $D = \varepsilon(r)E$ between the displacement vectors $D$ and the electric field intensity $E$.

Keywords: electrostatics, solvation energy, smoothed dielectric permeability.

## INTRODUCTION

The calculation of the polar component $G_{solv}$ of the molecular solvation Gibbs energy is an important task; a great number of different methods have been developed for its solution [1–4]. Many of them use the construction of a molecular surface, within which the dielectric permittivity has one value (as a rule, close to unity) and another value outside it, for the calculation of $G_{solv}$ [3-7], These methods are characterized by speed and precision. But they also have many shortcomings. First, universal and exact algorithms for the construction of smooth molecular surfaces are very complicated and take a great deal of computer memory and time [5–7]. Second, a considerable transformation of a surface may occur for a small shift of atoms, thus resulting in nonphysical jumps of $G_{solv}$ during the structural optimization of a molecule in a solution. Moreover, the dielectric permittivity in real physical media changes continuously and smoothly rather than abruptly.

Another possible approach is to solve the Poisson–Boltzmann equation directly, dividing the entire space into cells, but this is associated with high computer memory and time costs [8]. A compact formulation of the method for the calculation of $G_{solv}$ in a medium with dielectric permittivity $\varepsilon(r)$, which is a differentiable function of the spatial coordinate, is given in the present work on the basis of a previously developed model [1, 2]. The proposed method allows the calculation of the solvation energy for a smoothly changing dielectric permittivity by means of an iterative process, where the integral is taken at each iteration only over the region within which a nonzero dielectric permittivity gradient exists. This allows us to appreciably reduce the expenses of the method in comparison with the direct solution of the Poisson–Boltzmann equation. The rapid convergence of the method already provides good results at the first iteration. The theoretical foundations of the method were laid in [9, 10]. A method with continuous dielectric permittivity was first proposed in [2]. In this work, the following solution was suggested:

$$\boldsymbol{E} = -\frac{\nabla \psi_0}{\varepsilon}. \tag{1}$$

where $\psi_0$ — is the electrostatic field that would be created by a molecule in the absence of a dielectric. However, this solution does not satisfy the Maxwell equations for electrostatics, as the electric field intensity curl is not equal to zero [10], i.e.,

$$\nabla \times \boldsymbol{E} = \frac{\nabla \varepsilon \times \nabla \psi_0}{\varepsilon^2} \neq 0,$$

if the vectors $\nabla \varepsilon$ and $\nabla \psi_0$ are not codirected. In actual fact, Eq. (1) is only the first approximation (however, sufficient in practice) for the electric field intensity, which is actually represented by the iterative series we obtained. The proposed method was further developed in [1], but the main emphasis in it is placed on the possibility of a nonlocal relationship between the displacement vectors and the electric field intensity.

This article is composed as follows. A new iterative scheme is described in Part I. The two models for the smoothing of a dielectric medium, viz., the model that was considered earlier in [2] and leads to gradient discontinuities and a new model, which allows us to perform infinite differentiation, are presented in Part II. The data in [2], which represent the results of the calculation of $G_{solv}$ for a certain set of molecules with consideration for only the first term of the expansion of Eq. (1) and their comparison with the experimental data and the calculation results obtained by the other method, are given in Part III as a validation of the method and illustrate the very rapid convergence of the proposed series. In support of this conclusion, the calculation of the two following expansion terms is given in the same part.

The polar Gibbs energy component calculated in the given work includes not only purely electrostatic energy but also both the component associated with the partial ordering of dipoles along a field (entropy component) and the nonelectrostatic energy of the "stretching" of dipoles along a field. However, no additional calculation is required for them, as the given components are already included in the energy calculated in this work. Since an insufficiently clear understanding of this fact is encountered, even in the serious literature [9], we thought it proper to clarify this question in the Appendix.

## I. ITERATION SCHEME OF THE METHOD.

Let a substrate molecule be placed in a solution. The charge distribution density inside this molecule is $\rho_0$, and the dielectric permittivity is equal to unity. The surrounding solution has the dielectric permittivity $\varepsilon_{out}$, but the transition at the boundary of a substrate is smoothed. Let us cite the basic equations of electrostatics:

$$\mathbf{D} = \varepsilon(\mathbf{r})\mathbf{E}, \quad (2)$$

$$\nabla \cdot \mathbf{D} = 4\pi\rho_0, \quad (3)$$

$$\mathbf{E} = -\nabla\psi. \quad (4)$$

where $\mathbf{E}$ - is the vector of the intensity of an electric field, $\mathbf{D}$ - is the vector of the displacement of an electric field, and $\psi$ - is the potential of an electric field. From the last equation it directly follows that $\nabla \times \mathbf{E} = 0$.

Under the influence of the electric field induced by the charges inside a molecule, the dielectric surrounding it acquires polarization, as a consequence of which a nonzero distribution of so-called bound charges occurs in the dielectric. Let us denote this distribution density as $\rho$. Then, the following relationship is satisfied:

$$\nabla \cdot \mathbf{E} = 4\pi(\rho_0 + \rho). \quad (5)$$

As is known, by knowing the density of the distribution of charges inside a molecule and the dielectric permittivity at each spatial point, we can univocally find the electrostatic field potential $\psi(\mathbf{r})$ at the boundary conditions

$$\lim_{r \to \infty} r\psi(\mathbf{r}) = const,$$

$$\lim_{r \to \infty} r^2 |\nabla\psi(\mathbf{r})| = const.$$

and, consequently, the intensity vector $\mathbf{E}$. Then, using Eq. (5), it is possible to find the bound charge distribution density $\rho$, from which we can express the energy of the interaction between a molecule and a solution.

The scheme for calculating the free salvation energy can be constructed with preservation of local relationship (1). Let us write the total potential in the Form

$$\psi = \frac{\psi_0}{\varepsilon} + \varphi, \quad (6)$$

where $\psi_0$ is the solution of the following equation:

$$\nabla^2 \psi_0 = -4\pi\rho_0, \quad (7)$$

i.e., it is the potential of the field that would be created in the absence of a dielectric. Then, for the field intensity $\mathbf{E}$ and the displacement $\mathbf{D}$ from Eqs. (2) and (4) we obtain

$$\mathbf{E} = -\frac{\nabla\psi_0}{\varepsilon} + \frac{\psi_0 \nabla\varepsilon}{\varepsilon^2} - \nabla\varphi,$$

$$\mathbf{D} = -\nabla\psi_0 + \frac{\psi_0 \nabla\varepsilon}{\varepsilon} - \varepsilon\nabla\varphi. \quad (8)$$

With consideration for Eqs. (7) and (8), condition (3) can be rewritten as

$$\nabla\left(\frac{\psi_0 \nabla \varepsilon}{\varepsilon} - \varepsilon \nabla \varphi\right) = 0. \tag{9}$$

Since the divergence of the expression within the brackets is equal to zero, this expression is the curl of a certain vector field $B$:

$$\nabla \times B = \frac{\psi_0 \nabla \varepsilon}{\varepsilon} - \varepsilon \nabla \varphi. \tag{10}$$

Whence we can express $\nabla \varphi = \frac{\psi_0 \nabla \varepsilon}{\varepsilon^2} - \frac{\nabla \times B}{\varepsilon}$ and substitute it into Eq. (8):

$$E = -\frac{\nabla \psi_0}{\varepsilon} + \frac{\nabla \times B}{\varepsilon}.$$

(11)

Then the intensity curl equal to zero can be expressed via the vector field $B$ as:

$$\frac{\nabla \varepsilon \times \nabla \psi_0}{\varepsilon} + \varepsilon \nabla \times \left(\frac{\nabla \times B}{\varepsilon}\right) = \frac{\nabla \varepsilon \times \nabla \psi_0}{\varepsilon} - \frac{\nabla \varepsilon \times (\nabla \times B)}{\varepsilon} + \nabla \times (\nabla \times B) = 0. \tag{12}$$

The latter equality represents the equation for the vector field $B$. Let us rewrite it in the form of a linear integral equation as:

$$\nabla \times B = \frac{1}{4\pi} \int \frac{\left[-\frac{\nabla' \varepsilon \times \nabla' \psi_0}{\varepsilon} + \frac{\nabla' \varepsilon \times (\nabla' \times B)}{\varepsilon}, r - r'\right]}{|r - r'|^3} d^3r' =$$

$$= \frac{1}{4\pi} \int \left[\nabla \frac{1}{|r - r'|}, \left(-\frac{\nabla' \varepsilon \times \nabla' \psi_0}{\varepsilon} + \frac{\nabla' \varepsilon \times (\nabla' \times B)}{\varepsilon}\right)\right] d^3r'. \tag{13}$$

One possible method of solving Eq. (13) is associated with the representation of $B$ as the sum $B = B_1 + B_2 + ...$, in which the curl of each following summand is appreciably lower than the curl of the preceding summand.
In this case, $\nabla \varepsilon$ and $\nabla \times B_1$ are considered to have the same (first) order of smallness. Based on the assumption that $\nabla \times B$ is small in comparison with $\nabla \psi_0$, the first term of the series can be selected as

$$\nabla \times B_1 = \frac{1}{4\pi} \int \frac{\left[-\frac{\nabla' \varepsilon \times \nabla' \psi_0}{\varepsilon}, r - r'\right]}{|r - r'|^3} d^3r' = \frac{1}{4\pi} \int \left[\nabla \frac{1}{|r - r'|}, -\frac{\nabla' \varepsilon \times \nabla' \psi_0}{\varepsilon}\right] d^3r'.$$

(14)

Substituting Eq. (14) into Eq. (12) and preserving the terms of the same order of smallness, we obtain the equation for the following term of the series:

$$-\frac{\nabla \varepsilon \times (\nabla \times B_1)}{\varepsilon} + \nabla \times (\nabla \times B_2) = 0. \tag{15}$$

Whence we can express the second term: as

$$\nabla \times B_2 = \frac{1}{4\pi} \int \frac{\left[\frac{\nabla' \varepsilon \times (\nabla' \times B_1)}{\varepsilon}, r - r'\right]}{|r - r'|^3} d^3r' = \frac{1}{4\pi} \int \left[\nabla \frac{1}{|r - r'|}, \frac{\nabla' \varepsilon \times (\nabla' \times B_1)}{\varepsilon}\right] d^3r'. \tag{16}$$

Hence, we obtain the following recurrent relationship for the terms of the series:

$$\nabla \times \boldsymbol{B}_{n+1} = \frac{1}{4\pi} \int \frac{\left[\frac{\nabla'\varepsilon \times (\nabla' \times \boldsymbol{B}_n)}{\varepsilon}, \boldsymbol{r}-\boldsymbol{r}'\right]}{|\boldsymbol{r}-\boldsymbol{r}'|^3} d^3\boldsymbol{r}'. \qquad (17)$$

Using this relationship, we can find the vector $\nabla \times \boldsymbol{B}$. Taking Eq. (11) into consideration, it is possible to express the density of bound charges via this vector from Eq. (5) as

$$\rho = \frac{1}{4\pi}\left(\frac{\nabla \psi_0 \cdot \nabla \varepsilon}{\varepsilon^2} + \nabla \cdot \left(\frac{\nabla \times \boldsymbol{B}}{\varepsilon}\right)\right) = \frac{1}{4\pi}\frac{\nabla \varepsilon}{\varepsilon^2} \cdot (\nabla \psi_0 - (\nabla \times \boldsymbol{B})) \qquad (18)$$

The polar component of the free solvation energy can be calculated by the following formula

$$G_{solv} = \frac{1}{2}\int \frac{\rho_0(\boldsymbol{r})\rho(\boldsymbol{r}')}{|\boldsymbol{r}-\boldsymbol{r}'|} d^3r d^3r'. \qquad (19)$$

In the approximation $\nabla \times \boldsymbol{B} = 0$, we have the following expression for the solvation energy:

$$G_{solv}^{(1)} = \frac{1}{8\pi}\int \frac{\rho_0(\boldsymbol{r})}{|\boldsymbol{r}-\boldsymbol{r}'|} \frac{\nabla'\psi_0 \cdot \nabla'\varepsilon}{[\varepsilon(\boldsymbol{r}')]^2} d^3r d^3r'. \qquad (20)$$

As was already mentioned above, this first expansion term coincides with the approximation proposed in [2]. Moreover, this term was also described in [1].

The other series terms determined from Eqs.(14) and (17) are small, if
    (1) The dielectric permittivity gradient itself is small; and
    (2) The field created by the charges inside a substrate is close to the dielectric permittivity gradient by its direction.

When the width of the smooth transition of the dielectric permittivity tends to zero at the boundary of a substrate, the series we obtained becomes equivalent to the iterative series of the polarizable continuum model (PCM) method [3].

## II. MODEL OF THE SMOOTH CHANGE IN THE DIELECTRIC PERMITTIVITY AT THE SUBSTRATE–SOLVENT BOUNDARY

Let us describe how the dielectric permittivity must change to satisfy the condition of the smallness of its gradient, if it is equal to unity inside a molecule and to $1 + 4\pi\{\chi_\infty + \chi_{in}\}$, where $\chi_\infty$ and $\chi_{in}$ − are the positive constants that are responsible for the fast (electronic) and slow (nuclear) dielectric permittivities in a solvent. The calculation of the dielectric permittivity near a substrate was based on the continuum model developed in [2]

$$\varepsilon(\mathbf{r}) = 1 + 4\pi z(\mathbf{r})\{\chi_\infty + \chi_{in}\varphi(z(\mathbf{r}))\}, \tag{21}$$

where $z(\mathbf{r})$ – is a certain function, which is univocally determined by the distance from a spatial point $\mathbf{r}$ to the surface of a molecule and described below and $\varphi(\mathbf{r}) = exp\left\{\frac{(z-1)}{z_0}\right\}$ ($z_0$ – is a positive constant) (Fig. 2). In this case, the dielectric permittivity smoothly changes from unity inside the molecule to $1 + 4\pi(\chi_\infty + \chi_{in})$ in the solvent.

Let us now specify the behavior of the function $z(\mathbf{r})$. Let the molecular surface be specified by van der Waals spheres with the radii $R_j$ and their centers in the atoms, whose radius vectors are $\mathbf{R}_j$; $j=1,..., N$ (the current number of an atom). Let $l(\mathbf{r})$ – be the distance from the point $\vec{r}$ to the molecular surface. If the point $\mathbf{r}$ does not lie inside one of the atoms, i.e., if $\|\mathbf{r} - \mathbf{R}_j\| > R_j$ for all j, $l(\mathbf{r}) = min_j \, l_j(\mathbf{r})$, where

$$l_j(\mathbf{r}) = \|\mathbf{r} - \mathbf{R}_j\| - R_j,$$
otherwise $l(\mathbf{r}) = 0$. \tag{22}

In [2] the following formula is proposed for $z(\mathbf{r})$:

$$z(\mathbf{r}) = 1 + exp\left\{-\frac{2l(\mathbf{r})}{\delta}\right\} - 2\, exp\left\{-\frac{l(\mathbf{r})}{\delta}\right\}, \tag{23}$$

where $\delta$ – is a small positive constant, which has the meaning of the transition width. However, the choice $z(\mathbf{r})$ is bad because the derivative of this function has discontinuities at the points where the distances to two or more atoms are equal.

For faster convergence of the method, in this work it is recommended to use another smooth function $z(\mathbf{r})$, which is infinitely differentiable at any point:

$$z(\mathbf{r}) = \frac{\sum_j exp\left\{-\frac{\delta}{l_j(\mathbf{r})}\right\}\left(1 - exp\left\{-\frac{\delta_1^2}{(l_j(\mathbf{r}) - l(\mathbf{r}))^2}\right\}\right)}{\sum_j \left(1 - exp\left\{-\frac{\delta_1^2}{(l_j(\mathbf{r}) - l(\mathbf{r}))^2}\right\}\right)}, \tag{24}$$

with the constants $\delta$ and $\delta_1$ satisfying the condition $\delta_1 \ll \delta < R_j$.

**III. COMPARISON OF THE RESULTS OF CALCULATING THE POLAR ENERGY BY THE METHOD PROPOSED IN THIS WORK WITH THE EXPERIMENTAL DATA AND THE RESULTS OBTAINED BY THE POLARIZABLE CONTINUUM MODEL METHOD.**

The numerical calculations of the solvation energy were performed in [2] for a number of molecules in the approximation $\nabla \times \mathbf{B} = 0$ and compared with the results obtained by the PCM method [3] and experimental data. The dielectric permittivity was calculated by Eqs. (21)–(23) with the following parameters:

$z_0 = 0.5$,
$\delta = 0.32\ A$,
$\varepsilon_{out} = 1 + 4\pi\{\chi_\infty + \chi_{in}\} = 78$,
$\varepsilon_\infty = 1 + 4\pi\chi_\infty = 1.77$,

The charges of atoms were taken from the quantum-mechanical calculations by the Hartree–Fock method (6-31G(d,p) базис) according to the electrostatic potential (ESP) based scheme for the localization of a charge [11]. The values of radii were taken from Table 5 of [12] and multiplied by $k = 0.92$ for the proposed method and by $k = 1.06$ for the PCM method.

The experimental data were taken from [12,13]. To separate out the polar component from the experimental Gibbs energy of the solvation of an ionized molecule, we subtracted the experimental Gibbs energy of the solvation of a nonionized molecule with a similar structure; the difference between the nonpolar components was close to zero. The obtained result was compared with the corresponding difference between the calculated polar components of the Gibbs energy of the solvation of these molecules. The comparison of the polar Gibbs energies (see figure) indicates the rapid convergence of the considered method, as the first term of the series we obtained already provides a sufficient accuracy of the model.

Since the question of how strong the contribution of additional expansion terms is may arise, we performed the numerical calculation of the first and the two following expansion terms ($G_{solv}^{(1)}$ from (20)) from Eqs. (14), (18), and (19), and ($G_{solv}^{(2)}$ из формул (14), (18), (19), $G_{solv}^{(3)}$ from Eqs. (16), (18) and (19)) respectively) for the five ions that had already been considered earlier in [2]. The corresponding integrals need to be numerically calculated only within a narrow region near the surface of a molecule. Outside this region, the subintegral function is close to zero, owing to the small value of the gradient ε. The integrals have good mathematical convergence at the discontinuity points of the subintegral function. These facts considerably facilitate the task of numerical integration. In contrast to [2], the charges of atoms were calculated by the Merck molecular force field 94 (MMFF94) model. Correspondingly, the radii from Table 5 of [12] were multiplied by $k = 0.8$. The other parameters were taken to be the same as in [2]. The calculation results in kcal/mol are given in the table. From the given results it can be seen that the total correction introduced by the two following expansion terms is very small and has the same order of magnitude as the root-mean-square deviation $G_{solv}^{(1)}$ of the calculated $\Delta G_{эксп}$ from its experimental values (1.75 kcal/mol). The results of the calculation of the first correction $G_{solv}^{(1)}$ in Fig. 3 are taken from [2] and were very close to those obtained in this work.

## CONCLUSIONS

In this work, a compact formulation of the iterative method for calculating the polar component of the solvation Gibbs energy $G_{solv}$ was proposed for the case of the continuous distribution of the dielectric permittivity in a solution. Such a distribution gives a more physically valid model of a solvent at its boundary with a substrate, prevents $G_{solv}$ jumps upon the shift of atoms, and allows us to easily calculate the cases with a smooth change in the dielectric permittivity of a solvent. A smooth infinitely differentiable model was proposed for the dielectric permittivity at the substrate–solvent boundary. The comparison of the numerical calculation for the first iteration $G_{solv}^{(1)}$ already gives good agreement with the experimental data and the values $G_{solv}$ obtained by the PCM method.

From the results of the calculation of the two following expansion terms it can be seen that their sum is small and has the same order of magnitude as the root-mean-square deviation of the first correction from its experimental value.

**ACKNOWLEDGMENTS**
We are very grateful to M.V. Basilevsky for providing the materials that were used in this work.

**APPENDIX**

The question of what value, the potential electrostatic internal energy or the Gibbs energy, is calculated with the use of the dielectric permittivity often arises [9]. Below, we give a consideration of this question for a model system to confirm the latter.

Let us first consider this question qualitatively. The Gibbs energy can be written as

$$G_{free} = U_{el} + U_{nonel} - TS + pV, \tag{25}$$

where $U_{el}$ – is the potential electrostatic energy

$$U_{el} = \int \frac{E^2}{8\pi} dV, \tag{26}$$

is the nonelectrostatic energy, i.e., the potential energy of the tension of "springs"; it is present when a dipole in a dielectric is described in the form of two charges coupled by a spring and absent for hard rotating dipoles,

$TS$ is the entropy component, where the entropy is decreased in the process of solvation owing to the ordering of dipoles upon the polarization of a medium, and

$pV$ is an insubstantial component, as the volume does not almost change upon solvation. The polar component of the solvation energy is determined by the formula

$$G_{pol} = \int \frac{\varepsilon E^2}{8\pi} dV. \tag{27}$$

From the fact that it does not coincide with $U_{el}$ from Eq. (26), it is already evident that it cannot be the potential electrostatic energy. On the other hand, it can be illustrated qualitatively that $G_{pol}$ has the properties of free energy, i.e., upon solvation

$$\Delta G_{free} = \Delta G_{pol}. \tag{28}$$

For T = 0 and hard dipoles in the process of solvation, we have from Eq. (25)

$$\Delta G_{free} = \Delta U_{el} = \int \frac{E_2^2}{8\pi} dV - \int \frac{E_1^2}{8\pi} dV \tag{29}$$

and from Eq. (27)

$$\Delta G_{pol} = \Delta U_{el} = \int \frac{\varepsilon E_2^2}{8\pi} dV - \int \frac{E_1^2}{8\pi} dV. \tag{30}$$

If Eq. (28) is valid, this is possible only for ε = 1. Actually, at absolute zero all hard dipoles are completely oriented in a single direction and the linear response of the polarization of a medium to a field is absent, whence ε = 1, as expected from Eqs. (29) and (30).

For T = 0 and dipoles on "springs" in the process of solvation, we have from Eq. (25)

$$\Delta G_{free} = \Delta U_{el} + \Delta U_{nonel} = \int \frac{E_2^2}{8\pi} dV - \int \frac{E_1^2}{8\pi} dV + \Delta U_{nonel} \tag{31}$$

and from Eq. (27)

$$\Delta G_{pol} = \Delta U_{el} = \int \frac{\varepsilon E_2^2}{8\pi} dV - \int \frac{E_1^2}{8\pi} dV . \qquad (32)$$

If Eq. (28) is valid, we may expect a nonzero response of the polarization of a medium $\varepsilon > 1$ due to the $\Delta U_{nonel}$. Actually, nothing prevents such dipoles from being stretched at absolute zero, producing the linear response of the polarization of a medium to a field, i.e., $\varepsilon > 1$, as expected from Eqs. (31) and (32).

Let us describe the situation for hard dipoles qualitatively. We shall consider the change in the free energy of the system of dipoles placed in a total (intrinsic + external) electric field with the intensity E. Let us assume that the field is weakly changed on the magnitude of a dipole. By the magnitude of a dipole its geometric size is meant. Then, for the free energy change $\Delta G$, we can write

$$\Delta G = G(E) - G(0) = RT \ln \int \exp(-\beta(\sum - d_i E(r_i) \cos \vartheta_i) d\Gamma \qquad (33)$$

The integration in Eq. (33) is performed over the entire phase space,, $d_i$ is the $i$th dipole momentum modulus, $E(r_i)$ - is the total field intensity at the center of the $i$th dipole momentum, $\cos \vartheta_i$ is the angle between the external field and the dipole momentum, and the summation in the exponent is conducted for all the dipoles. Note that the assumption on the absence of the interaction between induced dipole momenta is not used in Eq. (33). Let us take the variation of Eq. (33) over the function of the external field intensity. We obtain

$$\frac{\delta(\Delta G)}{\delta E} = \frac{\int \sum - d_i \cos \vartheta_i \exp(-\beta(\sum - d_i E(r_i) \cos \vartheta_i)) d\Gamma}{\int \exp(-\beta(\sum - d_i E(r_i) \cos \vartheta_i) d\Gamma} . \qquad (34)$$

It can be seen that Eq. (34) is simply the average value of the projection of an induced dipole momentum onto the direction of the external field intensity. Then, we obtain

$$\frac{\delta(\Delta G)}{\delta E} = p = (\varepsilon - 1)\varepsilon_0 E \qquad (35)$$

Here, we have used the fact that the induced dipole momentum $p$ of the unit volume is related to the dielectric susceptibility $\chi$ as

$$p = \chi \varepsilon_0 E . \qquad (36)$$

and the dielectric permittivity is related to the dielectric susceptibility as

$$\varepsilon = 1 + \chi . \qquad (37)$$

Integrating Eq. (35), we obtain

$$\Delta G = \frac{(\varepsilon - 1)\varepsilon_0 E^2}{2} + C. \tag{38}$$

The integration constant is determined from the condition $\Delta G(\varepsilon = 1) = 0$, whence we obtain C = 0. Finally, we have

$$\Delta G = \frac{(\varepsilon - 1)\varepsilon_0 E^2}{2}, \tag{39}$$

as we set out to prove.

Table 1. Calculated and experimental solvation energies (in kcal/mol)

| Ион | $\Delta G_{экс п}$[2] | $G_{solv}^{(1)}$ | $G_{solv}^{(2)}$ | $G_{solv}^{(3)}$ | $\Delta G_{solv}=G_{solv}^{(2)}+G_{solv}^{(3)}$ |
|---|---|---|---|---|---|
| $(CH_3)_2NH_2^+$ | -65,9 | -65,89 | -3,22 | 1,67 | -1,55 |
| $(CH_3)_3NH^+$ | -58,9 | -60,05 | -2,33 | 0,66 | -1,67 |
| $CH_3(CH_2)_2NH_3^+$ | -68,8 | -71,30 | -8,61 | 6,01 | -2,60 |
| $CH_3CH_2NH_3^+$ | -70,4 | -72,57 | -6,27 | 5,49 | -0.78 |
| $CH_3NH_3^+$ | -73,1 | -75,23 | -3,03 | 2,02 | -1,01 |

Список рисунков.
Fig. 1. Finding distance to molecule. From [2].
Fig. 2. Behavior of function *z( l )*. From [2].
Fig. 3. Correlation of (a) $G_{solv}^{(1)}$ и $G_{\exp}$ with a root_mean_square deviation of 1.75 and (b) $G_{solv}^{(1)}$ и $G_{PCM}$ with a root_mean_square deviation of 1.68 kcal/mol for ammonium type ions. Here $G_{solv}^{(1)}$ is the solvation Gibbs energy obtained within the framework of only a single (first) iteration, $G_{PCM}$ is the solvation Gibbs energy determined by the PCM model, and $G_{\exp}$ - is the solvation Gibbs energy calculated from experimental data. All energy values are given in kcal/mol. The data from [2] are used.

Рис. 1.

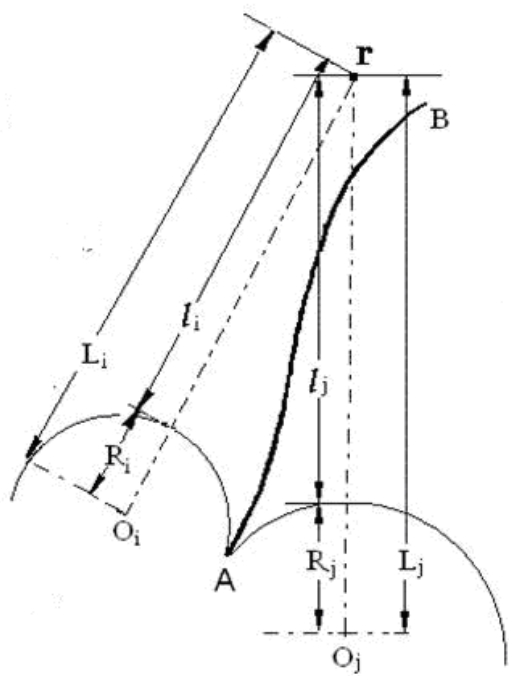

Рис. 2.

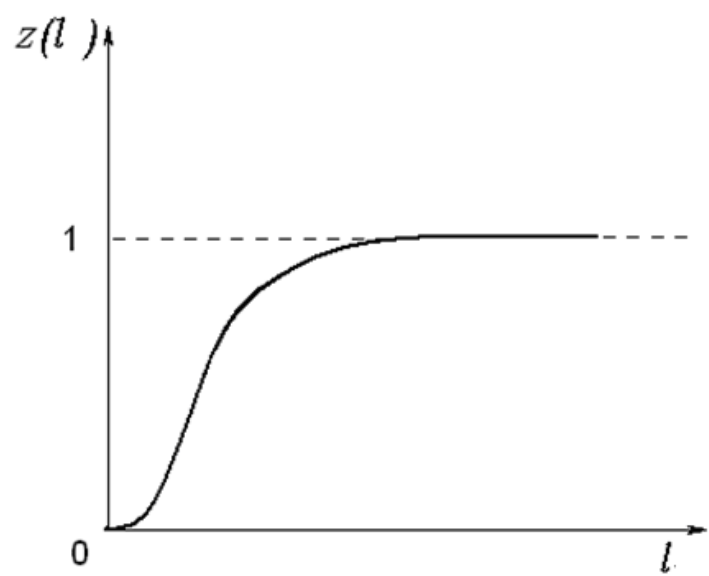

Рис.3.

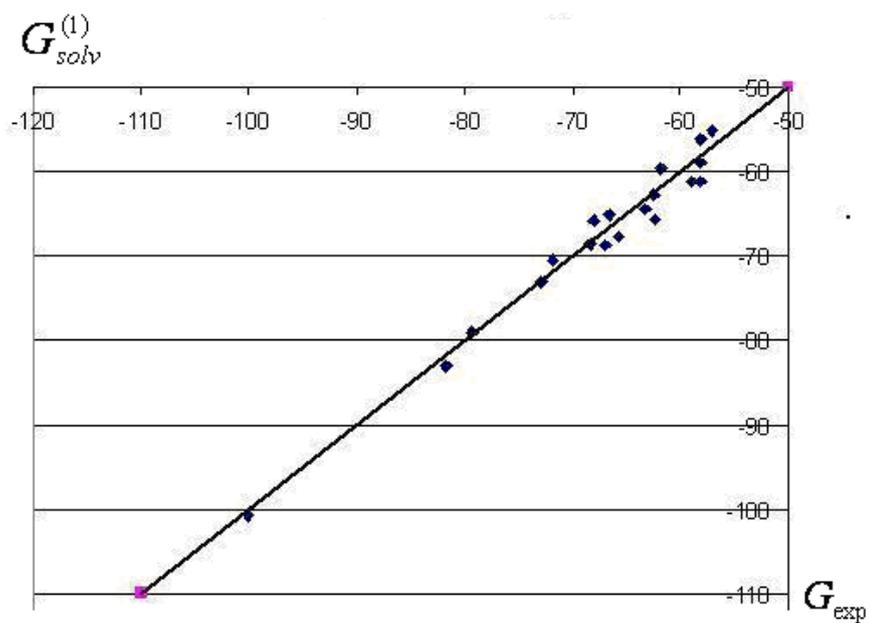

a)

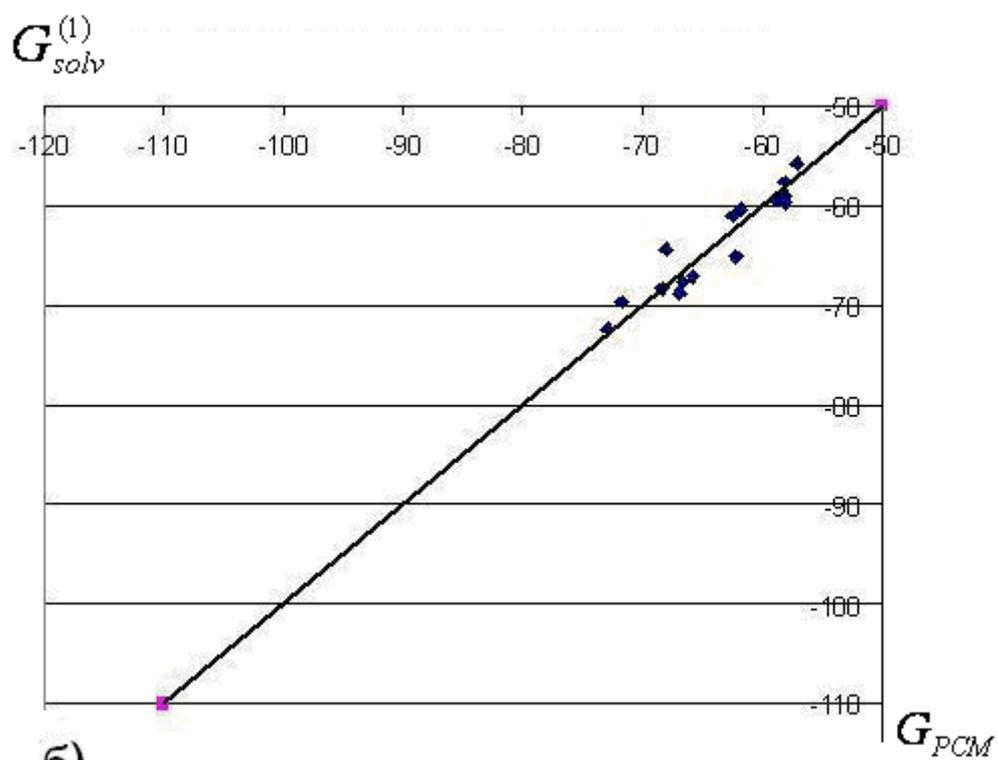

б)

# ИТЕРАЦИОННЫЙ МЕТОД РАСЧЕТА ПОЛЯРНОЙ СОСТАВЛЯЮЩЕЙ ЭНЕРГИИ ГИББСА РАСТВОРЕНИЯ МОЛЕКУЛ ПРИ УСЛОВИИ ПЛАВНОГО ИЗМЕНЕНИЯ ДИЭЛЕКТРИЧЕСКОЙ ПРОНИЦАЕМОСТИ РАСТВОРА

Григорьев Ф.В.[1*], Купервассер О.Ю.[1**], Кикоть И.П.[2***]

[1]*Научно-исследовательский вычислительный центр Московского государственного университета им.М.В. Ломоносова*

*Учреждение Российской академии наук Институт химической физики им. Н.Н.Семенова Российской академии наук, Москва*

*E-mail:* fedor.grigoriev@gmail.com,

**E-mail:* olegkup@yahoo.com,

***E-mail:* irakikotx@gmail.com

На основе ранее развитой модели формулируется итерационный метод расчета полярной составляющей энергии Гиббса растворения, предполагающий плавное изменение диэлектрической проницаемости как между субстратом и растворителем, так и внутри растворителя. Метод развит в приближении локальной связи между векторами смещения $D$ и напряженности электрического поля $E$: $D = \varepsilon(r)E$.

Ключевые слова: электростатика, энергия сольватации, сглаженная диэлектрическая проницаемость.

1. **Введение**

Расчет полярной компоненты энергии Гиббса сольватации молекулы $G_{solv}$ представляет собой важную задачу, для решения которой было создано большое число различных методов [1-4]. Многие из них используют для расчета $G_{solv}$ построение молекулярной поверхности [3-7], внутри которой диэлектрическая проницаемость имеет одно значение (как правило, близкое к единице), а вовне — другое. Эти методы отличает быстрота и точность. Но они имеют и много недостатков: во-первых, универсальные и точные алгоритмы построения гладких молекулярных поверхностей очень сложны и занимают много компьютерной памяти и времени [5-7]. Во-вторых, при малом сдвиге атомов может происходить значительная перестройка поверхности, что приводит к нефизическим скачкам $G_{solv}$ при оптимизации структуры молекулы в растворе. Кроме того, в реальных физических средах изменение диэлектрической проницаемости происходит непрерывно и гладко, а не скачком.

Другой возможный подход состоит в том, чтобы напрямую решать уравнение Пуассона-Больцмана, разбивая все пространство на ячейки, но это связанно с большими затратами компьютерной памяти и времени [8] В настоящей работе на основе развитой ранее модели [1-2] дается компактная формулировка метода для расчета $G_{solv}$ в среде с диэлектрической проницаемостью $\varepsilon(r)$, которая является дифференцируемой функцией координат пространства. Предлагаемый метод позволяет рассчитывать энергию сольватации для плавно меняющейся диэлектрической проницаемости путем итерационного процесса, где на каждой итерации берется интеграл лишь по области, где существует ненулевой градиент диэлектрической проницаемости. Это позволяет значительно снизить затраты метода по сравнению с прямым решением уравнения Пуассона-Больцмана. Быстрота сходимости метода обеспечивает хорошие результаты уже для первой итерации.

Теоретические основы метода были заложены в работах [9-10]. Впервые метод с непрерывной диэлектрической проницаемостью был предложен в [2]. В этой работе предлагалось такое решение:

$$\boldsymbol{E} = -\frac{\nabla \psi_0}{\varepsilon}. \tag{1}$$

где $\psi_0$ — электростатическое поле, которое создавалось бы молекулой в отсутствие диэлектрика. Однако это решение не удовлетворяет уравнениям Максвелла для электростатики - ротор напряженности электрического поля не равен нулю [10]:

$\nabla \times \boldsymbol{E} = \dfrac{\nabla \varepsilon \times \nabla \psi_0}{\varepsilon^2} \neq 0$ , если векторы $\nabla \varepsilon$ и $\nabla \psi_0$ не сонаправлены.

На самом деле выражение (1) является лишь *первым* приближением (достаточным, впрочем, на практике) для напряженности электрического поля, которая в действительности представляется полученным нами итерационным рядом. Дальнейшее развитие предложенный метод получил в работе [1], однако основной акцент в ней сделан на возможность нелокальной связи между векторами смещения и напряженности электрического поля.

Статья построена следующим образом. Вначале приводится описание новой итерационной схемы. Затем приводятся две модели сглаживания ди-электрической среды – одна уже ранее рассмотренная в работе [2], приводящая к разрывам в градиенте, а другая новая, позволяющая проводить бесконечное дифференцирование. В качестве доказательства работоспособности метода в третьей части приведены данные из работы [2] - результаты расчетов $G_{solv}$ для некоторого набора молекул с учетом только первого члена полученного разложения (1) и их сравнение с результатами эксперимента и расчетов другим методом, которые иллюстрируют очень быструю сходимость предлагаемого ряда. Также дается расчет двух следующий членов разложения, подтверждающий этот вывод.

Полярная компонента энергии Гиббса, которая ищется в данной работе, включает не только чисто электростатическую энергию: в нее также входят как компонента, связанная с частичным упорядочиванием диполей вдоль поля (энтропийная составляющая), так и неэлектростатическая энергия «растяжения» диполей вдоль поля. Однако никакого их дополнительного расчета не требуется – данные компоненты уже включены в вычисленную в статье энергию. Поскольку даже в серьезной литературе [9] встречается недостаточно ясное понимание этого факта, мы сочли нужным добавить Приложение, проясняющее этот вопрос.

## 2. Итерационная схема метода.

Пусть имеется молекула субстрата, помещенного в раствор. Плотность распределения заряда внутри этой молекулы $\rho_0$, а диэлектрическая проницаемость равна 1. Окружающий раствор имеет диэлектрическую проницаемость $\varepsilon_{out}$, но переход на границе субстрата сглаживается. Основные уравнения электростатики:

$$\boldsymbol{D} = \varepsilon(\boldsymbol{r})\boldsymbol{E}, \tag{2}$$

$$\nabla \cdot \boldsymbol{D} = 4\pi\rho_0, \tag{3}$$

$$\boldsymbol{E} = -\nabla \psi. \tag{4}$$

где $\boldsymbol{E}$ - вектор напряженности электрического поля, $\boldsymbol{D}$ - вектор смещения электрического поля, $\psi$ - потенциал электрического поля. Из последнего уравнения непосредственно следует, что $\nabla \times \boldsymbol{E} = 0$.

Под влиянием электрического поля зарядов внутри молекулы происходит поляризация окружающего ее диэлектрика, вследствие чего возникает ненулевое распределение так называемых связанных зарядов в диэлектрике. Обозначим эту плотность распределения $\rho$. Тогда выполняется следующее соотношение:

$$\nabla \cdot \boldsymbol{E} = 4\pi(\rho_0 + \rho). \tag{5}$$

Как известно, зная распределение зарядов внутри молекулы $\rho$ и диэлектрическую проницаемость в каждой точке пространства, можно однозначно найти потенциал электрического поля $\psi(\boldsymbol{r})$ при граничных условиях

$$\lim_{r \to \infty} r\psi(\boldsymbol{r}) = const,$$
$$\lim_{r \to \infty} r^2 |\nabla \psi(\boldsymbol{r})| = const.$$

а, следовательно, и вектор напряженности $E$. Из уравнения (5) тогда можно найти плотность связанных зарядов $\rho$, а через плотность выразить энергию взаимодействия молекулы и раствора.

Схему для расчета свободной энергии сольватации можно построить, сохранив локальную связь (1). Запишем полный потенциал в виде:

$$\psi = \frac{\psi_0}{\varepsilon} + \varphi, \qquad (6)$$

где $\psi_0$ является решением следующего уравнения

$$\nabla^2 \psi_0 = -4\pi\rho_0, \qquad (7)$$

то есть является потенциалом поля, которое создавалось бы в отсутствие диэлектрика. Тогда для напряженности поля $E$ и смещения $D$ из уравнений (4) и (2) получим:

$$\begin{aligned} E &= -\frac{\nabla \psi_0}{\varepsilon} + \frac{\psi_0 \nabla \varepsilon}{\varepsilon^2} - \nabla \varphi, \\ D &= -\nabla \psi_0 + \frac{\psi_0 \nabla \varepsilon}{\varepsilon} - \varepsilon \nabla \varphi. \end{aligned} \qquad (8)$$

С учетом (7) и (8) условие (3) можно переписать так:

$$\nabla \left( \frac{\psi_0 \nabla \varepsilon}{\varepsilon} - \varepsilon \nabla \varphi \right) = 0. \qquad (9)$$

Поскольку дивергенция выражения в скобках равна 0, то это выражение является ротором некоторого векторного поля $B$:

$$\nabla \times B = \frac{\psi_0 \nabla \varepsilon}{\varepsilon} - \varepsilon \nabla \varphi. \qquad (10)$$

Отсюда можно выразить $\nabla \varphi = \frac{\psi_0 \nabla \varepsilon}{\varepsilon^2} - \frac{\nabla \times B}{\varepsilon}$ и подставить в уравнение (8):

$$E = -\frac{\nabla \psi_0}{\varepsilon} + \frac{\nabla \times B}{\varepsilon}. \qquad (11)$$

Тогда ротор напряженности, который равен нулю, может быть следующим образом выражен через векторное поле $B$:

$$\frac{\nabla \varepsilon \times \nabla \psi_0}{\varepsilon} + \varepsilon \nabla \times \left( \frac{\nabla \times B}{\varepsilon} \right) = \frac{\nabla \varepsilon \times \nabla \psi_0}{\varepsilon} - \frac{\nabla \varepsilon \times (\nabla \times B)}{\varepsilon} + \nabla \times (\nabla \times B) = 0. \qquad (12)$$

Последнее равенство представляет собой уравнение на векторное поле $\boldsymbol{B}$. Перепишем его в виде линейного интегрального уравнения:

$$\nabla \times \boldsymbol{B} = \frac{1}{4\pi} \int \frac{\left[ -\frac{\nabla'\varepsilon \times \nabla'\psi_0}{\varepsilon} + \frac{\nabla'\varepsilon \times (\nabla' \times \boldsymbol{B})}{\varepsilon}, \boldsymbol{r} - \boldsymbol{r}' \right]}{|\boldsymbol{r} - \boldsymbol{r}'|^3} d^3\boldsymbol{r}' = $$
$$= \frac{1}{4\pi} \int \left[ \nabla \frac{1}{|\boldsymbol{r} - \boldsymbol{r}'|}, \left( -\frac{\nabla'\varepsilon \times \nabla'\psi_0}{\varepsilon} + \frac{\nabla'\varepsilon \times (\nabla' \times \boldsymbol{B})}{\varepsilon} \right) \right] d^3\boldsymbol{r}'. \qquad (13)$$

Один из возможных путей решения (13) связан с представлением $\boldsymbol{B}$ в виде суммы $\boldsymbol{B} = \boldsymbol{B}_1 + \boldsymbol{B}_2 + \ldots$, в которой ротор каждого последующего слагаемого существенно меньше ротора предыдущего. При этом считается, что $\nabla\varepsilon$ и $\nabla \times \boldsymbol{B}_1$ имеют одинаковый (первый) порядок малости. Исходя из предположения, что $\nabla \times \boldsymbol{B}$ мало по сравнению с $\nabla\psi_0$, первый член ряда может быть выбран следующим образом:

$$\nabla \times \boldsymbol{B}_1 = \frac{1}{4\pi} \int \frac{\left[ -\frac{\nabla'\varepsilon \times \nabla'\psi_0}{\varepsilon}, \boldsymbol{r} - \boldsymbol{r}' \right]}{|\boldsymbol{r} - \boldsymbol{r}'|^3} d^3\boldsymbol{r}' = \frac{1}{4\pi} \int \left[ \nabla \frac{1}{|\boldsymbol{r} - \boldsymbol{r}'|}, -\frac{\nabla'\varepsilon \times \nabla'\psi_0}{\varepsilon} \right] d^3\boldsymbol{r}'. \qquad (14)$$

Подставляя (14) в (12) и сохраняя члены одного порядка малости, получим уравнение для следующего члена ряда:

$$-\frac{\nabla\varepsilon \times (\nabla \times \boldsymbol{B}_1)}{\varepsilon} + \nabla \times (\nabla \times \boldsymbol{B}_2) = 0.$$

(15)

Отсюда можно выразить второй член:

$$\nabla \times \boldsymbol{B}_2 = \frac{1}{4\pi} \int \frac{\left[ \frac{\nabla'\varepsilon \times (\nabla' \times \boldsymbol{B}_1)}{\varepsilon}, \boldsymbol{r} - \boldsymbol{r}' \right]}{|\boldsymbol{r} - \boldsymbol{r}'|^3} d^3\boldsymbol{r}' = \frac{1}{4\pi} \int \left[ \nabla \frac{1}{|\boldsymbol{r} - \boldsymbol{r}'|}, \frac{\nabla'\varepsilon \times (\nabla' \times \boldsymbol{B}_1)}{\varepsilon} \right] d^3\boldsymbol{r}'. \qquad (16)$$

Таким образом, получаем следующее рекуррентное соотношение для членов ряда:

$$\nabla \times \boldsymbol{B}_{n+1} = \frac{1}{4\pi} \int \frac{\left[ \frac{\nabla'\varepsilon \times (\nabla' \times \boldsymbol{B}_n)}{\varepsilon}, \boldsymbol{r} - \boldsymbol{r}' \right]}{|\boldsymbol{r} - \boldsymbol{r}'|^3} d^3\boldsymbol{r}'. \qquad (17)$$

Используя это соотношение, можно найти вектор $\nabla \times \boldsymbol{B}$. С учетом (11) можно теперь выразить плотность связанных зарядов через этот вектор из уравнения (5):

$$\rho = \frac{1}{4\pi}\left(\frac{\nabla \psi_0 \cdot \nabla \varepsilon}{\varepsilon^2} + \nabla \cdot \left(\frac{\nabla \times \boldsymbol{B}}{\varepsilon}\right)\right) = \frac{1}{4\pi}\frac{\nabla \varepsilon}{\varepsilon^2}\cdot(\nabla \psi_0 - (\nabla \times \boldsymbol{B})) \qquad (18)$$

Полярная составляющая свободной энергии растворения может вычислена по следующей формуле:

$$G_{solv} = \frac{1}{2}\int \frac{\rho_0(\boldsymbol{r})\rho(\boldsymbol{r}')}{|\boldsymbol{r}-\boldsymbol{r}'|} d^3r d^3r'. \qquad (19)$$

В приближении $\nabla \times \boldsymbol{B} = 0$ имеем следующее выражение для энергии растворения:

$$G_{solv}^{(1)} = \frac{1}{8\pi}\int \frac{\rho_0(\boldsymbol{r})}{|\boldsymbol{r}-\boldsymbol{r}'|}\frac{\nabla'\psi_0 \cdot \nabla'\varepsilon}{[\varepsilon(\boldsymbol{r}')]^2} d^3r d^3r'. \qquad (20)$$

Этот первый член разложения совпадает с предложенным в [2] приближением, как уже отмечалось выше. Кроме того, он описывается и в [1].

Другие члены ряда, определяемые из (14), (17) малы, если:
1) мал сам градиент диэлектрической проницаемости;
2) поле, создаваемое зарядами внутри субстрата, близко по направлению к градиенту диэлектрической проницаемости.

При стремлении ширины плавного перехода диэлектрической проницаемости к нулю на границе субстрата получаемый нами ряд становится эквивалентен итерационному ряду метода Polarizable Continuum Model [3].

### 3. Модель плавного изменения диэлектрической проницаемости на границе субстрата и растворителя.

Опишем, как должна меняться диэлектрическая проницаемость, чтобы выполнялось условие малости ее градиента, если известно, что внутри молекулы она равна 1, а в растворителе - $1 + 4\pi\{\chi_\infty + \chi_{in}\}$, где $\chi_\infty$, $\chi_{in}$ – константы, большие нуля, отвечающие за быструю (электронную) и медленную (ядерную) диэлектрические проницаемости. За основу расчета диэлектрической проницаемости около субстрата используем модель непрерывной среды, разработанной в [2]:

$$\varepsilon(\boldsymbol{r}) = 1 + 4\pi z(\boldsymbol{r})\{\chi_\infty + \chi_{in}\varphi(z(\boldsymbol{r}))\}, \qquad (21)$$

где $z(\boldsymbol{r})$ – некая функция, однозначно определяемая расстоянием от точки пространства $\boldsymbol{r}$ до поверхности молекулы, которая подробно описывается ниже, а $\varphi(\boldsymbol{r}) = exp\left\{\dfrac{(z-1)}{z_0}\right\}$ ($z_0$ – константа, большая нуля) (см. рис. 2). В этом случае диэлектрическая проницаемость плавно меняется от 1 внутри молекулы до $1 + 4\pi(\chi_\infty + \chi_{in})$ в растворителе.

Зададим теперь поведение функции $z(\boldsymbol{r})$. Пусть молекулярная поверхность задается сферами Ван-дер-ваальса с радиусами $R_j$ и центрами в атомах, радиус-векторы которых $\boldsymbol{R}_j$; $j=1,..., N$ – текущий номер атома. Пусть $l(\boldsymbol{r})$ – расстояние от точки $\vec{r}$ до молекулярной поверхности. Если точка $\boldsymbol{r}$ не лежит внутри одного из атомов, т.е. если $\|\boldsymbol{r} - \boldsymbol{R}_j\| > R_j$ для всех j, тогда $l(\boldsymbol{r}) = min_j\, l_j(\boldsymbol{r})$, где $l_j(\boldsymbol{r}) = \|\boldsymbol{r} - \boldsymbol{R}_j\| - R_j$,

иначе $l(\boldsymbol{r}) = 0$. \hfill (22)

В работе [2] предлагается следующая формула для $z(\boldsymbol{r})$:

$$z(\boldsymbol{r}) = 1 + exp\left\{-\dfrac{2l(\boldsymbol{r})}{\delta}\right\} - 2\,exp\left\{-\dfrac{l(\boldsymbol{r})}{\delta}\right\}, \qquad (23)$$

где $\delta$ – малая константа, большая нуля, имеющая смысл ширины перехода. Однако такой выбор $z(\boldsymbol{r})$ плох тем, что производная этой функции имеет разрыв в точках, где расстояния до двух или более атомов одинаковы.

Для более быстрой сходимости метода в настоящей работе рекомендуется использовать другую гладкую функцию $z(\boldsymbol{r})$, которая является бесконечно дифференцируемой в любой точке:

$$z(r) = \frac{\sum_j exp\left\{-\frac{\delta}{l_j(r)}\right\}\left(1 - exp\left\{-\frac{\delta_1^2}{(l_j(r)-l(r))^2}\right\}\right)}{\sum_j \left(1 - exp\left\{-\frac{\delta_1^2}{(l_j(r)-l(r))^2}\right\}\right)}, \tag{24}$$

причем константы $\delta, \delta_1$ удовлетворяют условию $\delta_1 << \delta < R_j$.

## 4. Сравнение результатов расчета полярной энергии, проведенного предлагаемым в статье методом, с экспериментом и с результатами расчета методом PCM.

В работе [2] для приближения $\nabla \times \boldsymbol{B} = 0$ были проведены численные расчеты энергии растворения для ряда молекул, которые сравнивались с результатами расчета методом PCM [3] и экспериментальными данными. Для расчета диэлектрической проницаемости использовалась формулы (21-23) со следующими параметрами:

$z_0 = 0.5$,

$\delta = 0.32\,A$,

$\varepsilon_{out} = 1 + 4\pi\{\chi_\infty + \chi_{in}\} = 78$,

$\varepsilon_\infty = 1 + 4\pi\chi_\infty = 1.77$,

Заряды атомов были взяты из квантово-механических расчетов методом ХФ (6-31G(d,p) базис) согласно ESP схеме зарядовой локализации [13]. Радиусы были взяты из Таблицы 5 [12] и домножались на $k=0.92$ для предлагаемого метода и на $k=1.06$ для PCM. Экспериментальные данные приводятся из работ [11-12].

Для выделения полярной составляющей из экспериментальной величины энергии Гиббса растворения ионизированной молекулы вычиталась экспериментальная энергия Гиббса растворения аналогичной по структуре неионизированной молекулы, при этом разница неполярных составляющих близка к нулю. Полученный результат сравнивался с соответствующей разницей рассчитанных полярных компонент энергии Гиббса растворения этих двух молекул. Сравнение полярных энергий Гиббса (см. рис. 3) свидетельствуют о быстрой сходимости рассматриваемого метода – уже первый член полученного нами ряда обеспечивает достаточную точность модели.

Поскольку может возникнуть вопрос, насколько сильный вклад могут вносить дополнительные члены разложения, был проведен численный расчет первого ($G_{solv}^{(1)}$ из формулы (20)) и двух последующих членов разложения ($G_{solv}^{(2)}$ из формул (14,18,19), $G_{solv}^{(3)}$ из формул (16, 18, 19)) для пяти ионов, уже рассмотренных ранее в [2]. Соответствующие интегралы нужно численно подсчитывать лишь в узкой области вблизи поверхности молекулы. Вне этой области подынтегральная функция близка к нулю, из-за малой величины градиента ε. В точках разрыва подынтегральной функции интегралы имеют хорошую математическую сходимость. Эти факты значительно облегчают задачу численного интегрирования. Для расчета заряда атомов в отличие от [2] использовалась модель силового поля MMFF94. Соответственно радиусы из Таблицы 5 [12] домножались на *k=0.8*. Остальные параметры брались теми же, что и в [2]. Результат расчета (в ккал/моль) приводиться в Таблице 1.

Из приведенных результатов расчета можно видеть, что итоговая поправка, вносимая следующими двумя членами разложения, действительно мала и имеет тот же порядок величины, что и среднеквадратичное отклонение расчетов $G_{solv}^{(1)}$ от эксперимента $\Delta G_{эксп}$ (1.75 ккал/моль, см. подпись к рис. 3). Результаты расчета первой поправки $G_{solv}^{(1)}$ на рис. 3 взяты из [2] и очень близки по величине к полученным в данной работе.

## 5. Выводы.

В данной работе предложена компактная формулировка итерационного метода расчета полярной составляющей энергии Гиббса растворения $G_{solv}$ в случае непрерывного распределения диэлектрической проницаемости раствора. Подобное распределение дает более физически верную модель растворителя на границе с субстратом, предотвращает скачки $G_{solv}$ при сдвиге атомов, позволяет легко рассчитать ситуации с плавным изменением диэлектрической проницаемости растворителя. Предложена гладкая, бесконечно дифференцируемая модель для диэлектрической проницаемости на границе субстрата с растворителем. Сравнение численного расчета уже для первой итерации $G_{solv}^{(1)}$ дает хорошее совпадение с экспериментом и расчетом $G_{solv}$ на основе метода PCM.

Из результатов расчета двух следующих членов разложения можно видеть, что их сумма мала и имеет тот же порядок величины, что и среднеквадратичное отклонение первой поправки от эксперимента.

**Приложение**

Часто возникает вопрос о том, какая величина вычисляется с помощью диэлектрической проницаемости (полярная компонента энергии сольватации) – потенциальная электростатическая внутренняя энергия или энергия Гиббса [9]. В данном приложении содержится рассмотрение этого вопроса для модельной системы, подтверждающее последнее.

Рассмотрим вначале вопрос качественно. Энергию Гиббса можно записать так:

$$G_{free} = U_{el} + U_{nonel} - TS + pV, \qquad (25)$$

где $U_{el}$ – потенциальная электростатическая энергия:

$$U_{el} = \int \frac{E^2}{8\pi} dV, \qquad (26)$$

$U_{nonel}$ – неэлектростатическая энергия (т.е. потенциальная энергия натяжения «пружинок», которая присутствует, когда диполь в диэлектрике описывается как два заряда, соединенные пружинкой, и отсутствует для жестких вращающихся диполей),

$TS$ – энтропийная составляющая. Энтропия уменьшается при растворении, вследствие упорядочивания диполей при поляризации среды.

$pV$ – малосущественная компонента, поскольку объем почти не меняется при растворении.

Полярная компонента энергии сольватации определяется формулой:

$$G_{pol} = \int \frac{\varepsilon E^2}{8\pi} dV. \qquad (27)$$

Из того, что она не совпадает с $U_{el}$ (26), уже очевидно, что она не может быть потенциальной электростатической энергией. С другой стороны, можно проиллюстрировать качественно, что $G_{pol}$ обладает свойствами свободной энергии, т.е. при растворении

$$\Delta G_{free} = \Delta G_{pol}. \qquad (28)$$

Действительно, для Т=0 и жестких диполей при растворении из (25):

$$\Delta G_{free} = \Delta U_{el} = \int \frac{E_2^2}{8\pi}dV - \int \frac{E_1^2}{8\pi}dV \qquad (29)$$

и из (27)

$$\Delta G_{pol} = \Delta U_{el} = \int \frac{\varepsilon E_2^2}{8\pi}dV - \int \frac{E_1^2}{8\pi}dV \quad . \qquad (30)$$

Если $\Delta G_{free} = \Delta G_{pol}$ верно, то это возможно только для $\varepsilon=1$.

Действительно, жесткие диполи при абсолютном нуле все полностью ориентированы в одном направлении и линейный отклик поляризации среды на поле отсутствует – отсюда $\varepsilon=1$, как и ожидалось из (29) и (30)

Для T=0 и диполей на «пружинках» при растворении из (25):

$$\Delta G_{free} = \Delta U_{el} + \Delta U_{nonel} = \int \frac{E_2^2}{8\pi}dV - \int \frac{E_1^2}{8\pi}dV + \Delta U_{nonel} \qquad (31)$$

и из (27)

$$\Delta G_{pol} = \Delta U_{el} = \int \frac{\varepsilon E_2^2}{8\pi}dV - \int \frac{E_1^2}{8\pi}dV \quad . \qquad (32)$$

Если $\Delta G_{free} = \Delta G_{pol}$ верно, то ожидается ненулевой линейный отклик поляризации среды $\varepsilon>1$ за счет $\Delta U_{nonel}$. И действительно, ничто не мешает таким диполям растягиваться и при абсолютном нуле, давая линейный отклик поляризации среды на поле, т.е. $\varepsilon>1$ как и ожидалось из (31) и (32)

Опишем ситуацию для жестких диполей количественно. Рассмотрим изменение свободной энергии системы диполей, помещенных в полное электрическое поле (собственное + внешнее) напряженностью *E*. Предположим, что поле слабо меняется на величине диполя. Тогда можно записать для изменения свободной энергии $\Delta G$:

$$\Delta G = G(E) - G(0) = RT \ln \int \exp(-\beta(\sum -d_i E(r_i)\cos\vartheta_i)d\Gamma \quad . \qquad (33)$$

В (33) интегрирование ведется по всему фазовому пространству, $d_i$ модуль i-ого дипольного момента $E(r_i)$ - напряженность полного поля в центре i-ого дипольного момента, $\cos\vartheta_i$ - угол между внешнем полем и дипольным моментом, суммирование в экспоненте ведется по всем диполям. Отметим, что в (33) не используется предположение об отсутствии взаимодействия наведенных дипольных моментов.

Возьмем вариацию от (33) по функции напряженности внешнего поля. Получим

$$\frac{\delta(\Delta G)}{\delta E} = \frac{\int \sum -d_i \cos\vartheta_i \exp(-\beta(\sum -d_i E(r_i)\cos\vartheta_i))d\Gamma}{\int \exp(-\beta(\sum -d_i E(r_i)\cos\vartheta_i)d\Gamma} \quad . \tag{34}$$

Видно, что (34) – это просто среднее значение проекции индуцированного дипольного момента на направление напряженности внешнего поля. Тогда получаем

$$\frac{\delta(\Delta G)}{\delta E} = p = (\varepsilon-1)\varepsilon_0 E \tag{35}$$

Здесь мы использовали то, что индуцированный дипольный момент единицы объема $p$ связан с диэлектрической восприимчивостью $\chi$ как

$$p = \chi\varepsilon_0 E. \tag{36}$$

а диэлектрическая проницаемость с восприимчивостью как

$$\varepsilon = 1 + \chi. \tag{37}$$

Интегрируя уравнение (35), получим:

$$\Delta G = \frac{(\varepsilon-1)\varepsilon_0 E^2}{2} + C \tag{38}$$

Константа интегрирования $C$ определяется из условия $\Delta G(\varepsilon=1) = 0$, получаем $C=0$. Окончательно,

$$\Delta G = \frac{(\varepsilon-1)\varepsilon_0 E^2}{2}, \tag{39}$$

что и требовалось доказать.

# СПИСОК ЛИТЕРАТУРЫ

Таблица 1. Расчетные и экспериментальные энергии сольватации

| Ион | $\Delta G_{эксп}$[2] | $G_{solv}^{(1)}$ | $G_{solv}^{(2)}$ | $G_{solv}^{(3)}$ | $\Delta G_{solv}=G_{solv}^{(2)}+G_{solv}^{(3)}$ |
|---|---|---|---|---|---|
| $(CH_3)_2NH_2^+$ | -65,9 | -65,89 | -3,22 | 1,67 | -1,55 |
| $(CH_3)_3NH^+$ | -58,9 | -60,05 | -2,33 | 0,66 | -1,67 |
| $CH_3(CH_2)_2NH_3^+$ | -68,8 | -71,30 | -8,61 | 6,01 | -2,60 |
| $CH_3CH_2NH_3^+$ | -70,4 | -72,57 | -6,27 | 5,49 | -0.78 |
| $CH_3NH_3^+$ | -73,1 | -75,23 | -3,03 | 2,02 | -1,01 |

Список рисунков.

Рис. 1. Определение расстояния до молекулы. Из работы [2].

Рис. 2. Поведение функции *z( l )*. Из работы [2].

Рис. 3. Для ионов аммониевого типа корреляция величин $G_{solv}^{(1)}$ и $G_{\exp}$ ( а), среднеквадратичное отклонение 1.75 ккал/моль) и

$G_{solv}^{(1)}$ и $G_{PCM}$ (б, среднеквадратичное отклонение 1.68 ккал/моль); здесь $G_{solv}^{(1)}$ - энергии Гиббса растворения, полученная при расчете в рамках только одной (первой) итерации, $G_{PCM}$ - то же в рамках PCM, $G_{\exp}$ - пересчитана из эксперимента. Все энергетические величины даны в ккал/моль. Использованы данные из работы [2].

Рис. 1.

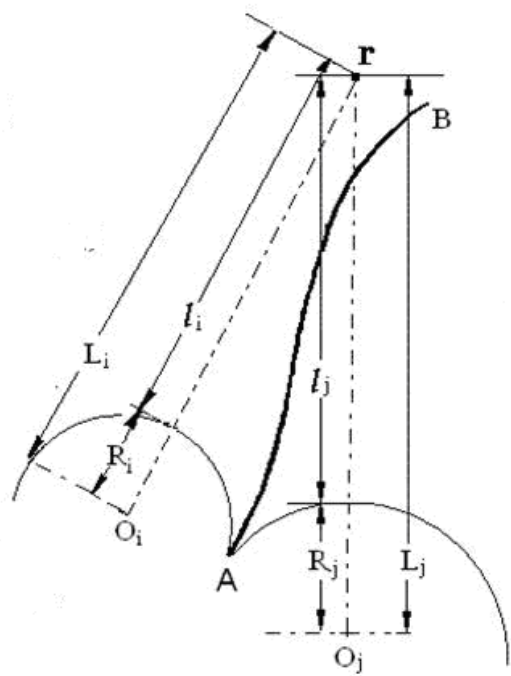

Рис. 2.

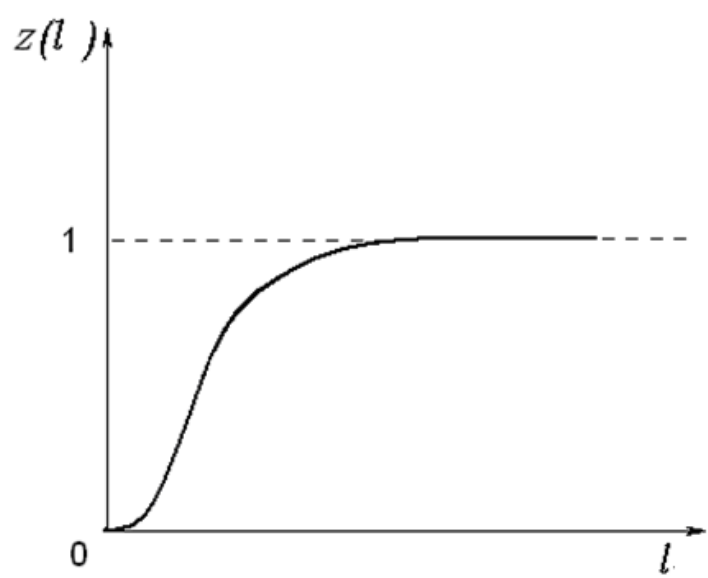

Рис.3.

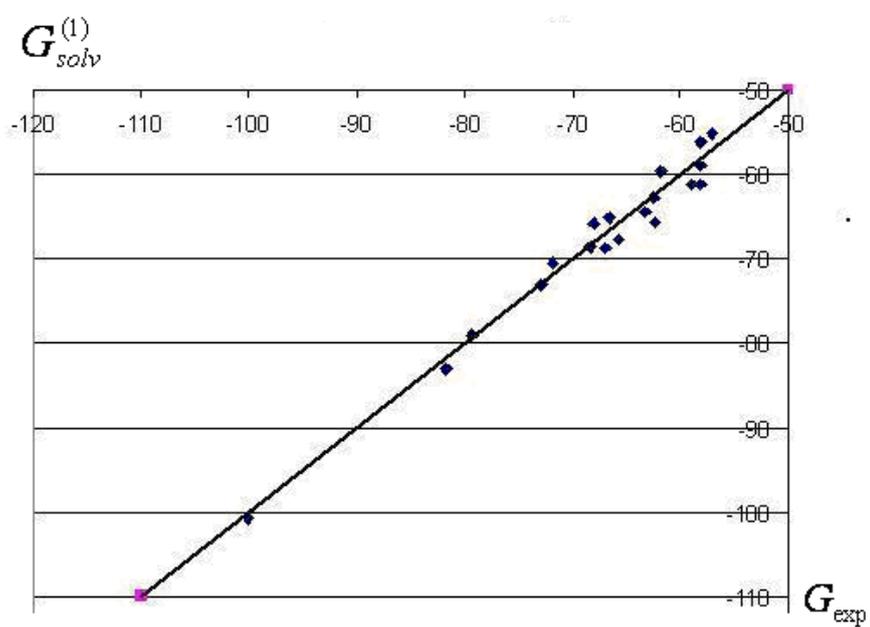

а)

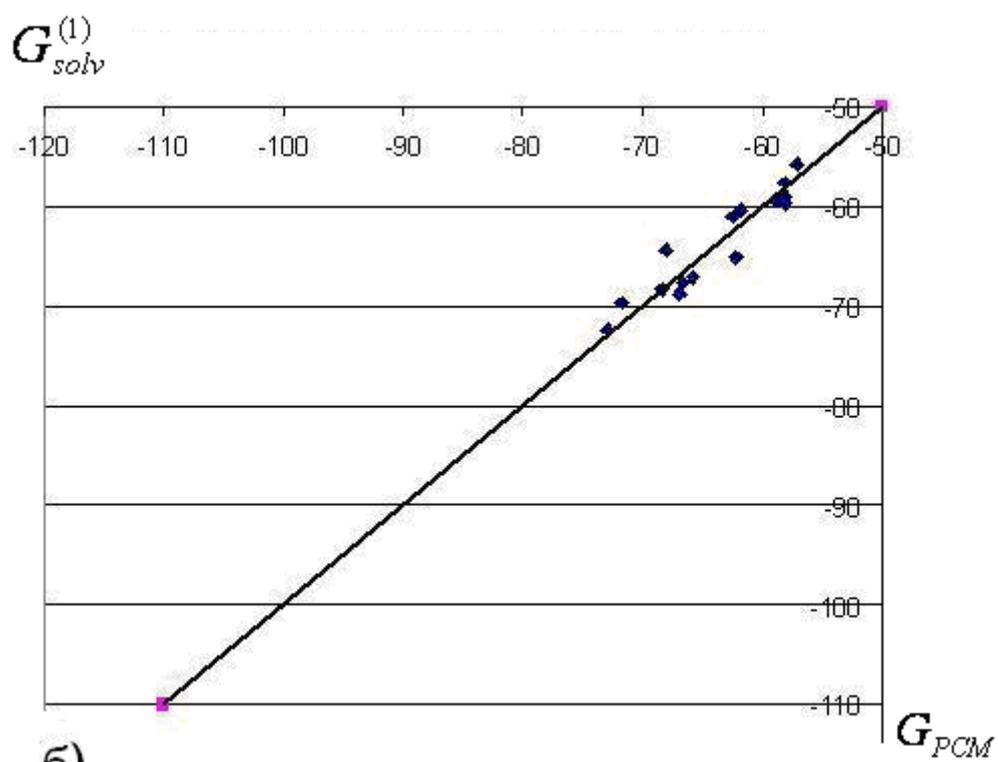

б)

20